\documentclass[amsmath,amssymb,amsfonts,aps,pre,preprint,superscriptaddress,bibnotes,showpacs,showkeys,longbibliography]{revtex4-1}

\usepackage{graphicx}
\usepackage{subfigure}
\usepackage{amsmath}
\usepackage{amsfonts}
\usepackage{amssymb}
\usepackage{natbib}
\usepackage{relsize}
\usepackage{sidecap}
\usepackage{braket}
\usepackage{footmisc}
\usepackage{footnote}
\usepackage{color}
\begin{document}

\title{Finite size effect on Bose-Einstein condensate mixtures in improved Hatree-Fock approximation}

\author{Nguyen Van Thu}
\affiliation{Department of Physics, Hanoi Pedagogical University 2, Hanoi, Vietnam}

\begin{abstract}
Using Cornwal-Jackiw-Tomboulis effective potential approach we found that at zero temperature, in improved Hatree-Fock approximation, the effective masses and order parameters of a two component Bose-Einstein condensates confined between two parallel plates strongly depend on the distance between two slabs. The Casimir force is also considered in this approximation and shown that this force differs from zero in limit of strong segregation.
\end{abstract}

\maketitle

\section{Introduction\label{sec:1}}

The original Casimir effect was discovered by H. B. G. Casimir \cite{Casimir}, which caused by the confinement of vacumm fluctuations of the electromagnetic field between two parallel plates at zero temperature. In this case the author pointed out that Casimir force is attractive and varying as a power $\ell^{-4}$ with $\ell$ being inter-distance beetwen two slabs. A review for Casimir effect and its applications were mentioned in \cite{Bordag}.

In field of single Bose-Einstein condensate (BEC), the Casimir-Polder force was measured in experiment by D. M. Harber {\it{et. al.}} in 2005 through center-of-mass oscillations of a BEC \cite{Harber}. In theoretical studies, using method of the quantum field theory in one-loop approximation, J. Schiefele and C. Henkel \cite{Schiefele} invoked Andersen's results \cite{Andersen} within framework of perturbative theory to consider the Casimir force of BEC at zero and finite temperature. Their results shown that the Casimir force is attractive and decays as the distance $\ell$ between two plates increases, which obeys the law $\ell^{-4}$.  However, their results could not give a general law because they only considered in the critical regions, where $\ell$ is larger/small enough. Using Euler-Maclurin formula, Biswas {\it{et. al.}} \cite{Biswas} obtained the analytical solution for Casimir force; therefore, we can estimate the distance-dependence of the Casimir force in detail. Employing the double parabola approximation proposed in \cite{Joseph} we considered both Casimir force, surface tension force and their combining \cite{Thunew}. At finite temperature, this effect was also investigated \cite{Dantchev,Biswas2}.

For two component Bose-Einstein condensates (BECs), in our previous paper \cite{Thu1}, the Casimir force was investigated in one-loop approximation. Several important results obtained, according to that, the Casimir force is not simple superposition of the one of two single component BEC and it is vanishing in some cases: (i) inter-distance between two plates becomes large enough; (ii) interaction is zero; (iii) interspecies interaction is full strong segregation. However, result (iii) is controversial because of explaining that the original Casimir force and interspecies interactive force are the same order in full strong separation. Developing these results, in this paper we research the finite size effect in a BECs in improved Hatree-Fock approximation (IHF) within Cornwal-Jackiw-Tomboulis (CJT) effective potential approach with the main aim is to find an answer for this question. Our system is confined to a parallel plate geometry with the size $\ell_x, \ell_y$ and inter-distance is $\ell = \ell_z$, which satisfies condition $\ell_x, \ell_y \gg \ell$. This means that our system is limited in the volume $V = \ell_x \ell_y \ell$ as was discussed in \cite{Lipowsky}.

This paper is organized as follow. In Section \ref{sec:2} we brief the CJT effective potential approach for BECs in infinite space.  The influence of finite size effect on effective masses, order parameters and Casimir force will be presented in Section \ref{sec:3}. The conclusions and outlook are  given in Section \ref{sec:4} to close the paper.

\section{A brief of CJT effective potential in improved Hatree-Fock approximation \label{sec:2}}

We start with a brief of CJT effective potential approach for a binary mixture of Bose gasses in double bubble approximation. Our system is described by the Lagrangian \cite{Pitaevskii,Pethick},
\begin{eqnarray}
{\cal L}=\sum_{j=1,2}\psi_j^*\left(-i\hbar\partial_t-\frac{\hbar^2}{2m_j}\nabla^2\right)\psi_j-V,\label{lagrangian}
\end{eqnarray}
with
\begin{eqnarray}
V=\sum_{j=1,2}\left(-\mu_j|\psi_j|^2+\frac{g_{jj}}{2}|\psi_j|^4\right)+g_{12}|\psi_1|^2|\psi_2|^2.\label{potentialGP}
\end{eqnarray}
Here $\mu_j$ and $m_j$ are chemical potential and atomic mass of component $j$, respectively. The coupling constants are given by $g_{jj}=4\pi\hbar^2a_{jj}/m_j>0$ and
\begin{eqnarray*}
g_{jj'}=2\pi\hbar^2\left(\frac{1}{m_j}+\frac{1}{m_{j'}}\right)a_{jj'}>0,
\end{eqnarray*}
with $a_{jj'}$ being the $s$-wave scattering length between components $j$ and $j'$. $\psi_j$ is the field operator.
Two condensates are immiscible \cite{AoChui}, that is when $g_{12}^2>g_{11}g_{22}$ and vice versa.

In tree approximation, by minimizing the tree potential we have gap equations
\begin{eqnarray}
-\mu_1\psi_1+g_{11}|\psi_1|^3+g_{12}\psi_1|\psi_2|^2&=&0,\nonumber\\
-\mu_2\psi_2+g_{22}|\psi_2|^3+g_{12}|\psi_1|^2\psi_22&=&0,\label{gaptree}
\end{eqnarray}
gives
\begin{eqnarray}
\psi_1^2=\frac{g_{22} \mu_1-g_{12}\mu_2}{g_{11} g_{22}-g_{12}^2},~\psi_2^2=\frac{g_{11} \mu_2-g_{12}\mu_1}{g_{11} g_{22}-g_{12}^2},\label{treesolution}
\end{eqnarray}
in broken phase.

We now focus on Hatree-Fock (HF) approximation. To begin with one first shifts the field operators
\begin{eqnarray}
\psi_j\rightarrow \psi_{j0}+\frac{1}{\sqrt{2}}(\psi_{j1}+i\psi_{j2}).\label{shift}
\end{eqnarray}
Plugging (\ref{shift}) into (\ref{lagrangian}) we get the interaction Lagrangian
\begin{eqnarray}
{\cal L}_{int}=&&\frac{1}{\sqrt{2}}\sum_{j=1,2}\left[g_{jj}\psi_{j0}\psi_{j1}+g_{12}\psi_{j'0}\psi_{j'1}\right](\psi_{j1}^2+\psi_{j2}^2)+\frac{1}{8}\sum_{j=1,2}g_{jj}(\psi_{j1}^2+\psi_{j2}^2)^2\nonumber\\
&&+\frac{g_{12}}{4}(\psi_{11}^2+\psi+{12}^2)(\psi_{21}^2+\psi_{22}^2).\label{Lint}
\end{eqnarray}
Combining (\ref{Lint}) and (\ref{treesolution}) we have the inverse propagator in tree approximation
\begin{eqnarray}
D_{j0}^{-1}=\left(
              \begin{array}{lr}
                \frac{\hbar^2k^2}{2m_j}+M_{j0}^2 & -\omega_n \\
                \omega_n & \frac{\hbar^2k^2}{2m_j} \\
              \end{array}
            \right),\label{protree}
\end{eqnarray}
in which $\omega_n=2\pi n T$ is Matsubara frequency and $T$ is temperature, $\vec{k}$ is wave vector. It is obvious that the excitations are phonon and with requirement
\begin{eqnarray*}
\det D_{j0}^{-1}=0,
\end{eqnarray*}
one obtains the dispersion relation
\begin{eqnarray}
E_{j}(k)=\sqrt{\frac{\hbar^2k^2}{2m_j}\left(\frac{\hbar^2k^2}{2m_j}+M_{j0}\right)},\label{disper0}
\end{eqnarray}
and the effective mass
\begin{eqnarray}
M_{10}^2&=&-\mu_1+3g_{11}\psi_{10}^2+g_{12}\psi_{20}^2,\nonumber\\
M_{20}^2&=&-\mu_2+3g_{22}\psi_{20}^2+g_{12}\psi_{10}^2.\label{Mj0}
\end{eqnarray}

Based on the interaction Lagrangian (\ref{Lint}) one has the effective potential $V_\beta^{CJT}$ at finite temperature in HF approximation
\begin{eqnarray}
V_{\beta}^{CJT}=&&\sum_{j=1,2}\left(-\mu_j|\psi_{j0}|^2+\frac{g_{jj}}{2}|\psi_{j0}|^4\right)+g_{12}|\psi_{10}|^2|\psi_{20}|^2\nonumber\\
&&+\frac{1}{2}\int_\beta\mbox{tr}\left\{\sum_{j=1,2}\left[\ln D_j^{-1}(k)+D_{j0}^{-1}(k)D(k)\right]-2.{1\!\!1}\right\}+\frac{3g_{11}}{8}(P_{11}^2+P_{22}^2)+\frac{g_{11}}{4}P_{11}P_{22}\nonumber\\
&&+\frac{3g_{22}}{8}(Q_{11}^2+Q_{22}^2)+\frac{g_{22}}{4}Q_{11}Q_{22}+\frac{g_{12}}{4}(P_{11}Q_{11}+P_{11}Q_{22}+P_{22}Q_{11}+P_{22}Q_{22}),\label{CJT1}
\end{eqnarray}
in which we abbreviate
\begin{eqnarray}
&&\int_\beta f(k)=T\sum_{n=-\infty}^{+\infty}\int \frac{d^3\vec{k}}{(2\pi)^3}f(\omega_n,\vec{k}),\nonumber\\
&&P_{aa}=\int_\beta P_{aa},~Q_{aa}=\int_\beta Q_{aa}.\label{momentum}
\end{eqnarray}
Based on many calculations, authors of Ref. \cite{Phat} proved that, the Goldstone theorem fails in the HF approximation. On purpose restoring this phonon we employ then method developed in \cite{Ivanov}. According to it, an extra term $\Delta V_\beta^{CJT}$ into the effective potential \ref{CJT1} and thus \cite{Phat},
\begin{eqnarray}
\widetilde{V}_\beta^{CJT}=&&\sum_{j=1,2}\left(-\mu_j|\psi_{j0}|^2+\frac{g_{jj}}{2}|\psi_{j0}|^4\right)+g_{12}|\psi_{10}|^2|\psi_{20}|^2\nonumber\\
&&+\frac{1}{2}\int_\beta\mbox{tr}\left\{\sum_{j=1,2}\left[\ln D_j^{-1}(k)+D_{j0}^{-1}(k)D(k)\right]-2.{1\!\!1}\right\}+\frac{g_{11}}{8}(P_{11}^2+P_{22}^2)+\frac{3g_{11}}{4}P_{11}P_{22}\nonumber\\
&&+\frac{g_{22}}{8}(Q_{11}^2+Q_{22}^2)+\frac{3g_{22}}{4}Q_{11}Q_{22}+\frac{g_{12}}{4}(P_{11}Q_{11}+P_{11}Q_{22}+P_{22}Q_{11}+P_{22}Q_{22}).\label{CJT2}
\end{eqnarray}
This approximation is called improved Hatree-Fock (IHF) approximation. From this effective potential:

- Minimizing this effective potential with respect to order parameters leads to gap equations
\begin{eqnarray}
-\mu_1+g_{11}\psi_{10}^2+g_{12}\psi_{20}^2+\Sigma_2^{(1)}&=&0,\nonumber\\
-\mu_2+g_{22}\psi_{20}^2+g_{12}\psi_{10}^2+\Sigma_2^{(2)}&=&0,\label{gap}
\end{eqnarray}
where
\begin{eqnarray}
\Sigma_2^{(1)}&=&\frac{1}{2}(3g_{11}P_{11}+g_{11}P_{22}+g_{12}Q_{11}+g_{12}Q_{22}),\nonumber\\
\Sigma_2^{(2)}&=&\frac{1}{2}(3g_{22}Q_{11}+g_{22}Q_{22}+g_{12}P_{11}+g_{12}P_{22}).\label{mass2}
\end{eqnarray}

- Minimizing this effective potential with respect to elements of the propagators one has Schwinger-Dyson (SD) equations \begin{eqnarray}
M_1^2&=&-\mu_1+3g_{11}\psi_{10}^2+g_{12}\psi_{20}^2+\Sigma_1^{(1)}\nonumber\\
M_2^2&=&-\mu_2+3g_{22}\psi_{20}^2+g_{12}\psi_{10}^2+\Sigma_1^{(2)}.\label{SD}
\end{eqnarray}
Here we use notations
\begin{eqnarray}
\Sigma_1^{(1)}&=&\frac{1}{2}(g_{11}P_{11}+3g_{11}P_{22}+g_{12}Q_{11}+g_{12}Q_{22}),\nonumber\\
\Sigma_1^{(2)}&=&\frac{1}{2}(g_{22}Q_{11}+3g_{22}Q_{22}+g_{12}P_{11}+g_{12}P_{22}).\label{mass}
\end{eqnarray}

- Combining the above, the propagators have the form
\begin{eqnarray}
D_{j}^{-1}=\left(
              \begin{array}{lr}
                \frac{\hbar^2k^2}{2m_j}+M_{j}^2 & -\omega_n \\
                \omega_n & \frac{\hbar^2k^2}{2m_j} \\
              \end{array}
            \right).\label{proIHF}
\end{eqnarray}
At this approximation, the Goldstone theorem is valid. Combining (\ref{SD}) and (\ref{gap}), the effective potential (\ref{CJT2}) reduces to
\begin{eqnarray}
\widetilde{V}_\beta^{CJT}=&&\sum_{j=1,2}\left(-\mu_j|\psi_{j0}|^2+\frac{g_{jj}}{2}|\psi_{j0}|^4\right)+g_{12}|\psi_{10}|^2|\psi_{20}|^2\nonumber\\
&&+\frac{1}{2}\int_\beta\mbox{tr}\left[\sum_{j=1,2}\ln D_j^{-1}(k)\right]-\frac{g_{11}}{8}(P_{11}^2+P_{22}^2)-\frac{3g_{11}}{4}P_{11}P_{22}\nonumber\\
&&-\frac{g_{22}}{8}(Q_{11}^2+Q_{22}^2)-\frac{3g_{22}}{4}Q_{11}Q_{22}-\frac{g_{12}}{4}(P_{11}Q_{11}+P_{11}Q_{22}+P_{22}Q_{11}+P_{22}Q_{22}),\label{CJT3}
\end{eqnarray}
and the dispersion relation in IHF approximation has the form
\begin{eqnarray}
E_j(k)=\sqrt{\frac{\hbar^2k^2}{2m_j}\left(\frac{\hbar^2k^2}{2m_j}+M_j^2\right)}.\label{dispIHF}
\end{eqnarray}

We now calculate the momentum integrals $P_{aa}$ and $Q_{aa}$. Using rules
\begin{eqnarray*}
\sum_{n=-\infty}^{+\infty}\frac{1}{\omega_n^2+E^2(k)}&=&\frac{1}{2TE(k)}\left[1+\frac{2}{e^{E(k)/k_BT}-1}\right],
\end{eqnarray*}
and combining (\ref{momentum}) with (\ref{proIHF}) one has at zero temperature
\begin{eqnarray}
&&P_{11}=\frac{1}{2}\int\frac{d^3\vec{k}}{(2\pi)^3}\sqrt{\frac{\hbar^2k^2/2m_1}{\hbar^2k^2/2m_1+M_1^2}},~P_{22}=\frac{1}{2}\int\frac{d^3\vec{k}}{(2\pi)^3}\sqrt{\frac{\hbar^2k^2/2m_1+M_1^2}{\hbar^2k^2/2m_1}},\nonumber\\
&&Q_{11}=\frac{1}{2}\int\frac{d^3\vec{k}}{(2\pi)^3}\sqrt{\frac{\hbar^2k^2/2m_2}{\hbar^2k^2/2m_2+M_2^2}},~Q_{22}=\frac{1}{2}\int\frac{d^3\vec{k}}{(2\pi)^3}\sqrt{\frac{\hbar^2k^2/2m_2+M_2^2}{\hbar^2k^2/2m_2}},\nonumber\\
&&\Omega_j\equiv\frac{1}{2}\int_\beta \mbox{tr}\ln D_j^{-1}(k)=\frac{1}{2}\int\frac{d^3\vec{k}}{(2\pi)^3}\sqrt{\frac{\hbar^2k^2}{2m_j}\left(\frac{\hbar^2k^2}{2m_j}+M_j^2\right)}.\label{tichphan}
\end{eqnarray}
In order to evaluate the integrals in (\ref{tichphan}) we introduce some dimensionless quantities $\kappa_j=k\xi_j$ with $\xi_j=\hbar/\sqrt{2m_jg_{jj}n_{j0}}$ being healing length, $n_{j0}$ is bulk density of component $j$. Effective mass is defined $\widetilde{M}_j^2=M^2_j/g_{jj}n_{j0}$. Based on these quantities, Eqs. (\ref{tichphan}) can be rewritten as
\begin{eqnarray}
&&P_{11}=\frac{1}{2\xi_1^3}\int\frac{d^3\kappa_1}{(2\pi)^3}\frac{\kappa_1}{\sqrt{\kappa_1^2+\widetilde{M}_1^2}},~P_{22}=\frac{1}{2\xi_1^3}\int\frac{d^3\kappa_1}{(2\pi)^3}\frac{\sqrt{\kappa_1^2+\widetilde{M}_1^2}}{\kappa_1},\nonumber\\
&&Q_{11}=\frac{1}{2\xi_2^3}\int\frac{d^3\kappa_2}{(2\pi)^3}\frac{\kappa_2}{\sqrt{\kappa_2^2+\widetilde{M}_2^2}},~Q_{22}=\frac{1}{2\xi_2^3}\int\frac{d^3\kappa_2}{(2\pi)^3}\frac{\sqrt{\kappa_2^2+\widetilde{M}_2^2}}{\kappa_2},\nonumber\\
&&\Omega_j=\frac{g_{jj}n_{j0}}{2\xi_j^3}\int\frac{d^3\kappa_j}{(2\pi)^3}\sqrt{\kappa_j^2(\kappa_j^2+\widetilde{M}_j^2)}.\label{tichphan1}
\end{eqnarray}

Using method of the dimensional regularization, above integrals are calculated \cite{Andersen},
\begin{eqnarray}
&&P_{11}=\frac{\widetilde{M}_1^3}{6\pi^2\xi_1^3},~P_{22}=-\frac{\widetilde{M}_1^3}{12\pi^2\xi_1^3},\nonumber\\
&&Q_{11}=\frac{\widetilde{M}_2^3}{6\pi^2\xi_2^3},~Q_{22}=-\frac{\widetilde{M}_2^3}{12\pi^2\xi_2^3}.\label{PQ1}
\end{eqnarray}
Plugging (\ref{PQ1}) into Eqs. (\ref{SD}) and (\ref{gap}) one has the gap and SD equations at zero temperature
\begin{eqnarray}
-1+\phi_j^2+K\phi_{j'}^2+\frac{5m_jg_{jj}\widetilde{M}_j^3}{12\pi^2\hbar^2\xi_j}+K\frac{m_{j'}g_{j'j'}\widetilde{M}_{j'}^3}{12\pi^2\hbar^2\xi_{j'}}&=&0,\nonumber\\
-1+3\phi_j^2+K\phi_{j'}^2-\frac{m_jg_{jj}\widetilde{M}_j^3}{12\pi^2\hbar^2\xi_j}+K\frac{m_{j'}g_{j'j'}\widetilde{M}_{j'}^3}{12\pi^2\hbar^2\xi_{j'}}&=&\widetilde{M}_j^2.\label{gapSD}
\end{eqnarray}

Note that we are considering here is at two-phase coexistence, this means that the pressures are the same for both components $P_1=P_2=P_0=g_{jj}n_{j0}^2/2$ . For a given system, solving numerically Eqs. (\ref{gapSD}) we obtain the effective masses and order parameters.

For free energy, using rule
\begin{eqnarray*}
T\sum_{n=-\infty}^{n=+\infty}\ln\left[\omega_n^2+E^2(k)\right]&=&E(k)+2T\ln\left[1-e^{-E(k)/k_BT}\right],
\end{eqnarray*}
one arrives
\begin{eqnarray}
\Omega_j\equiv\frac{1}{2}\int_\beta \mbox{tr}\ln D_j^{-1}(k)=\frac{1}{2}\int\frac{d^3\vec{k}}{(2\pi)^3}\sqrt{\frac{\hbar^2k^2}{2m_j}\left(\frac{\hbar^2k^2}{2m_j}+M_j^2\right)},\label{free}
\end{eqnarray}
at zero temperature. 

\section{Finite-size effect \label{sec:3}}

In this Section we investigate the influence of the finite size effect on our system. As mentioned above, our system is restricted between to parallel plates, which perpendiculars to $0z$-axis. These plates have large area and their distance is $\ell$. The Dirichlet boundary condition is applied at the plates.

\subsection{Effective masses and order parameters}

We first consider the effect from the compactified space along $z$-direction on effective masses and order parameters. Impose that the periodic boundary condition is applied, the wave vector is quantized as follows
\begin{eqnarray*}
k^2\rightarrow k_\perp^2+k_n^2, ~k_n=\frac{2n\pi}{\ell},
\end{eqnarray*}
or in dimensionless form
\begin{eqnarray}
\kappa_j^2\rightarrow \kappa_{j\perp}^2+\kappa_{jn}^2,~\kappa_{jn}=\frac{2\pi n}{L_j},\label{compac}
\end{eqnarray}
in which $L_j=\ell/\xi_j$. Under transformation (\ref{compac}), the momentum integrals (\ref{tichphan1}) have the form
\begin{eqnarray}
&&P_{11}=\frac{1}{2\xi_1^3}\sum_n \int \frac{d^2\kappa_1}{(2\pi)^2}\sqrt{\frac{\kappa_{1\perp}^2+(2\pi n/L_1)^2}{\kappa_{1\perp}^2+(2\pi n/L_1)^2+\widetilde{M}_1^2}},\nonumber\\
&&P_{22}=\frac{1}{2\xi_1^3}\sum_n \int \frac{d^2\kappa_1}{(2\pi)^2}\sqrt{\frac{\kappa_{1\perp}^2+(2\pi n/L_1)^2+\widetilde{M}_1^2}{\kappa_{1\perp}^2+(2\pi n/L_1)^2}}.\label{tichphan3}
\end{eqnarray}
In order to calculate (\ref{tichphan3}) one invokes Euler-Maclaurin formula \cite{Arfken},
 \begin{eqnarray}
\sum_{n=0}^\infty \theta_nF(n)-\int_0^\infty F(n)dn=-\frac{1}{12}F'(0)+\frac{1}{720}F'''(0)-\frac{1}{30240}F^{(5)}(0)+\cdots,\label{Euler}
\end{eqnarray}
and keeps up to thirst derivative term of (\ref{Euler}) and then takes a limit $\Lambda\rightarrow\infty$ for momentum cut-off leads
\begin{eqnarray}
&&P_{11}=-\frac{\pi^2m_1g_{11}n_{10}\xi_1^2}{90 \hbar ^2\ell^3 \widetilde{M}_1},~P_{22}=\frac{m_1g_{11}n_{10}\widetilde{M}_1}{12\hbar^2\ell}-\frac{m_1g_{11}n_{10}\xi_1^2\pi^2}{90\hbar^2\widetilde{M}_1\ell^3},\nonumber\\
&&Q_{11}=-\frac{\pi^2m_2g_{22}n_{20}\xi_2^2}{90 \hbar ^2\ell^3 \widetilde{M}_2},~Q_{22}=\frac{m_2g_{22}n_{20}\widetilde{M}_2}{12\hbar^2\ell}-\frac{m_2g_{22}n_{20}\xi_2^2\pi^2}{90\hbar^2\widetilde{M}_2\ell^3}.\label{tichphan4}
\end{eqnarray}

The system under consideration is in grand canonical ensemble therefore chemical potential is fixed $\mu_j=g_{jj}n_{j0}$. In equilibrium state the pressures are equal for both components $P_1=P_2=P_0=g_{jj}n_{j0}$.    Introducing $\phi_j=\psi_{j0}/\sqrt{n_{j0}}$, plugging (\ref{tichphan4}) into (\ref{gap}) yielding the gap equations
\begin{eqnarray}
&&-1+\phi_1^2+K\phi_2^2+\widetilde{\Sigma}_2^{(1)}=0,\nonumber\\
&&-1+\phi_2^2+K\phi_1^2++\widetilde{\Sigma}_2^{(2)}=0,\label{gap1}
\end{eqnarray}
in which
\begin{eqnarray*}
K=\frac{g_{12}}{\sqrt{g_{11}g_{22}}},
\end{eqnarray*}
and
\begin{eqnarray}
&&\widetilde{\Sigma}_2^{(1)}\approx \frac{m_1g_{11}\widetilde{M}_1}{24\hbar^2\ell}+K\frac{m_2g_{22}\widetilde{M}_2}{24\hbar^2\ell}-\frac{m_1g_{11}\xi_1^2\pi^2}{45\hbar^2\widetilde{M}_1\ell^3}-K\frac{m_2g_{22}\xi_2^2\pi^2}{90\hbar^2\widetilde{M}_2\ell^3},\nonumber\\
&&\widetilde{\Sigma}_2^{(1)}\approx\frac{m_2g_{22}\widetilde{M}_2}{24\hbar^2\ell}+K\frac{m_1g_{11}\widetilde{M}_1}{24\hbar^2\ell}-\frac{m_2g_{22}\xi_2^2\pi^2}{45\hbar^2\widetilde{M}_2\ell^3}-K\frac{m_1g_{11}\xi_1^2\pi^2}{90\hbar^2\widetilde{M}_1\ell^3}.\label{sig2}
\end{eqnarray}
Similarly, one has the SD equations
\begin{eqnarray}
&&\widetilde{M}_1^2=-1+3\phi_1^2+K\phi_2^2+\widetilde{\Sigma}_1^{(1)},\nonumber\\
&&\widetilde{M}_2^2=-1+3\phi_2^2+K\phi_1^2+\widetilde{\Sigma}_1^{(2)}.\label{SD1}
\end{eqnarray}
where
\begin{eqnarray}
&&\widetilde{\Sigma}_1^{(1)}=\frac{m_1g_{11}\widetilde{M}_1}{8\hbar^2\ell}+K\frac{m_2g_{22}\widetilde{M}_2}{24\hbar^2\ell}-\frac{m_1g_{11}\xi_1^2\pi^2}{45\hbar^2\widetilde{M}_1\ell^3}-K\frac{m_2g_{22}\xi_2^2\pi^2}{90\hbar^2\widetilde{M}_2\ell^3},\nonumber\\
&&\widetilde{\Sigma}_1^{(2)}=\frac{m_2g_{22}\widetilde{M}_2}{8\hbar^2\ell}+K\frac{m_1g_{11}\widetilde{M}_1}{24\hbar^2\ell}-\frac{m_2g_{22}\xi_2^2\pi^2}{45\hbar^2\widetilde{M}_2\ell^3}-K\frac{m_1g_{11}\xi_1^2\pi^2}{90\hbar^2\widetilde{M}_1\ell^3}.\label{sig1}
\end{eqnarray}

Mathematically, solving gap equations (\ref{gap1}) and SD equation (\ref{SD1}) one finds the $\ell$-dependence of effective masses $\widetilde{M}_j$ and order parameters $\phi_j$. These equations have no analytical solution, even have yet, it is no insight. In order to illustrate for these calculations, we are going to take the numerical computation for a binary mixture of Bose gases of rubidium in two different hyperfine states \cite{Egorov}. The first component BEC associated with $\left.\mid 1\right>=\left.\mid F=1,m_F=-1\right>$ and second one is $\left.\mid 2\right>=\left.\mid F=2,m_F=+1\right>$. For this system the parameters are in order $m_1=m_2=86.909u,~a_{11}=100.4a_0,a_{22}=95.44a_0,~\xi_1=4 \mu\mbox{m},~\xi_2=0.4 \mu\mbox{m}$. Here $u$ and $a_0$ are atomic mass unit and Bohr radius, respectively.
\begin{figure*}
  \mbox{
    \subfigure[\label{f1a}]{\includegraphics[scale=0.75]{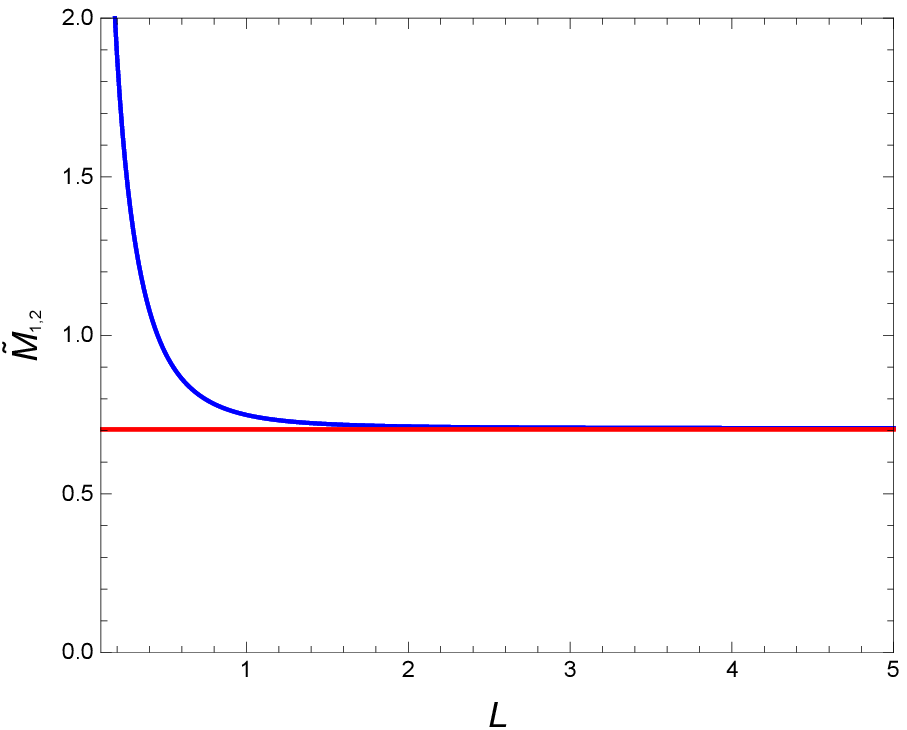}}\quad
    \subfigure[\label{f1b}]{\includegraphics[scale=0.75]{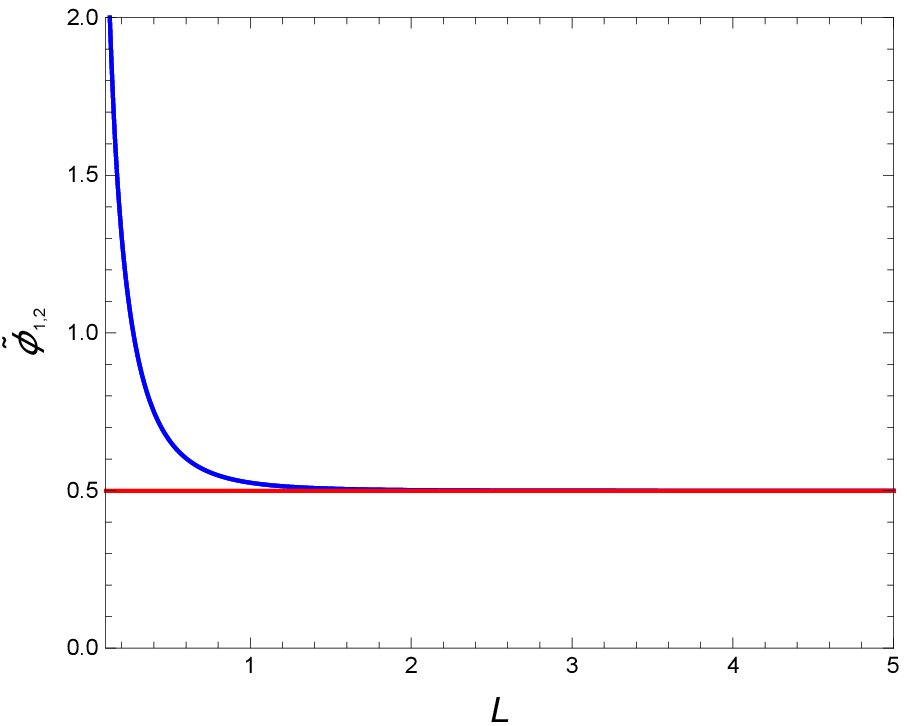}}
  }
  \caption{\footnotesize  (Color online) The effective masses (left) and order parameters (right) as a function of distance at $K=3$. The red and blue lines correspond to infinite and finite system.}
  \label{f1}
\end{figure*}

Figs. \ref{f1} we show the $L$-dependence of effective masses $M_{j}$ and order parameters $\phi_j$ at $K=3$. Both the effective masses and order parameters are divergent at $L=0$, decrease quite fast as $L$ increases and tend to constants when $L$ is large enough. The difference $\widetilde{M}_1-\widetilde{M}_1$ and $\phi_1-\phi_2$ are very small. For the infinite system, Eqs. (\ref{gapSD}) give $\widetilde{M}_1=0.7037,~\widetilde{M}_1=0.7038$ and $\phi_1=0.499071,~\phi_2=0.499035$. It is obvious that the finite size effect is significant in region $L\leq1$ at this value of $K$.

\begin{figure*}
  \mbox{
    \subfigure[\label{f1a}]{\includegraphics[scale=0.75]{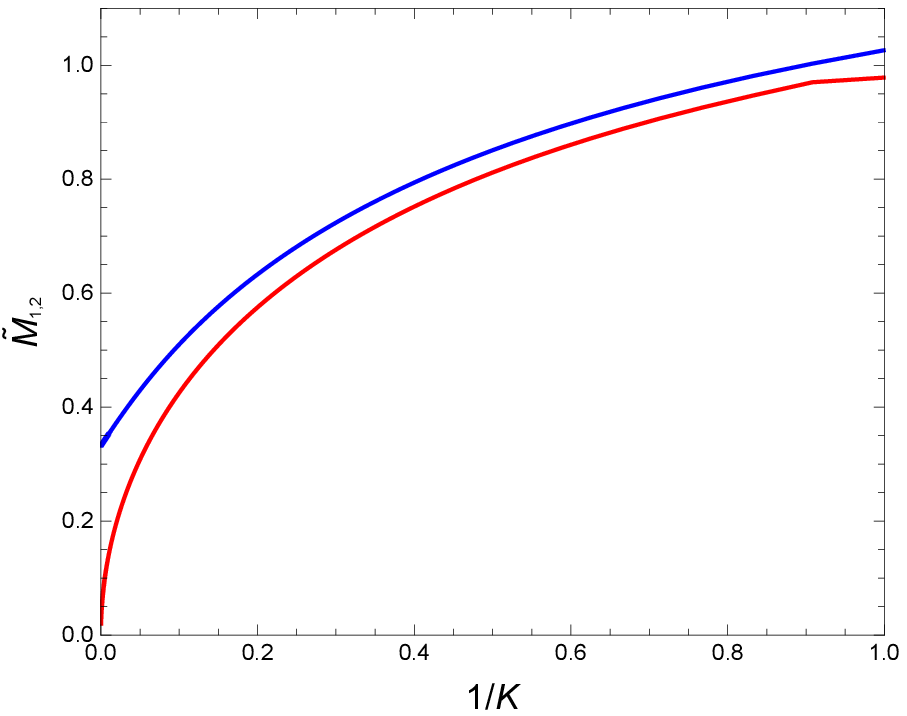}}\quad
    \subfigure[\label{f1b}]{\includegraphics[scale=0.75]{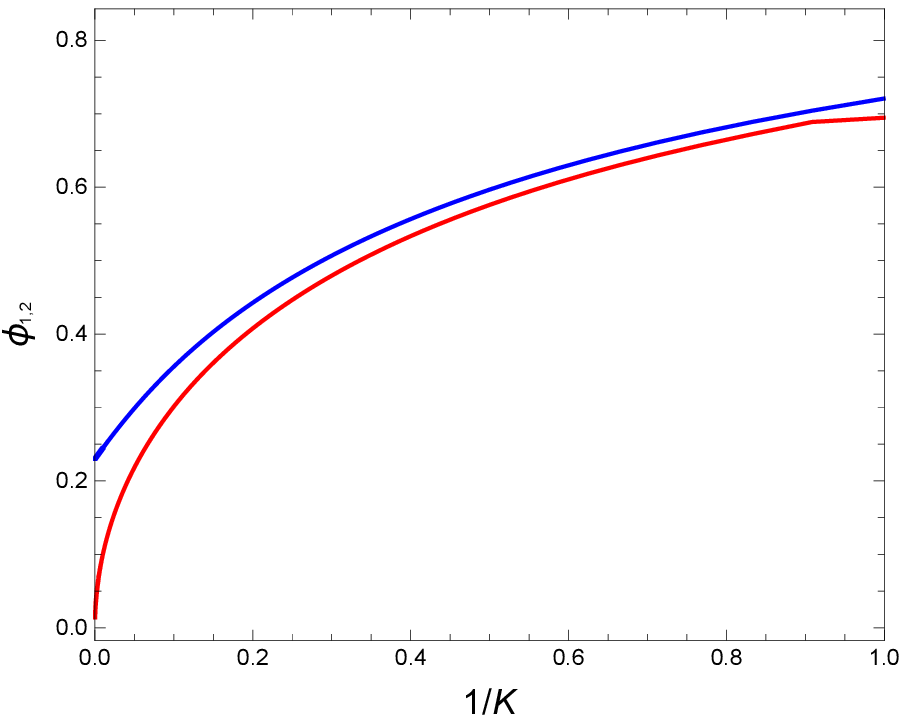}}
  }
  \caption{\footnotesize  (Color online) The effective masses (left) and order parameters (right) as a function of $K$ at $L=1$. The red and blue lines correspond to infinite and finite system.}
  \label{f2}
\end{figure*}

Similarly, the evolution of effective masses (left) and order parameters (right) versus $1/K$ are sketched in Fig. \ref{f2}. The blue lines correspond to those at $L=1$ and red lines associated with the infinite system. It shows that the finite size effect is clear. Especially, at $1/K \rightarrow 0$, i.e. strong segregated, the effective masses and order parameters are nonzero for finite system, whereas these quantities are vanishing for infinite system. Mathematically, taking a limit $K\rightarrow \infty$ for the gap and SD equations one has
\begin{eqnarray}
\widetilde{M}_j&=&\left(\frac{m_j g_{jj}\xi_j^2\pi^2}{45\hbar^2}\right)^{1/3}\ell^{-1},\nonumber\\
\phi_j&=&\frac{12 \sqrt[3]{5} \pi ^{4/3}g_{jj}^{2/3}m_j^{2/3} \xi_j^{4/3} \hbar ^{4/3}-(15 \pi )^{2/3}g_{jj}^{4/3}m_j^{4/3}\xi_j^{2/3}}{360 \sqrt[3]{3}\hbar ^{8/3}} \ell^{-2}.\label{strong1}
\end{eqnarray}
Eqs. (\ref{strong1}) confirm that both effective masses and order parameters diverge when $\ell$ tends to zero as shown in Figs. \ref{f1} for $K=3$.

In comparing to those in one-loop approximation mentioned in Ref. \cite{Thu1} one sees an important difference is that both effective masses and order parameters are independent on $\ell$ in one-loop approximation, whereas they depend strongly on distance in IHF approximation, especially in small-$\ell$ region. This is an improved results of IHF approximation.

From the above one can see that the finite size effect is significant for static quantities, for instance, the effective masses and order parameters. Based on these we can investigate the Casimir force in IHF approximation.

\subsection{Casimir force}

As already mentioned in many papers, for instance \cite{Casimir,Bordag} the Casimir effect at zero temperature caused by zero-point energy of a quantum field. In BEC(s) field, this effect associates with the quantum fluctuations on top of ground state, which corresponds to phononic excitations \cite{Thu1,Schiefele,Biswas,Biswas2,Biswas3}. We now consider it in IHF approximation and compare result with the one in one-loop approximation.

In order to calculate the Casimir force one first evaluate the free energy (\ref{free}). When the $z$-direction is compactified one has
\begin{eqnarray}
\Omega_j=\frac{g_{jj}n_{j0}}{2\xi_j^3}\sum_n \int \frac{d^2\kappa_j}{(2\pi)^2}\sqrt{\left[\kappa_{j\perp}^2+(2\pi n/L_j)^2\right]\left[\kappa_{j\perp}^2+(2\pi n/L_j)^2+\widetilde{M}_j^2\right]}.
\end{eqnarray}
Using Euler-Maclaurin formula (\ref{Euler}) as we did for momentum integrals, the free energy of BECs has the form
\begin{eqnarray}
\Omega&=&\sum_{j=1,2}\Omega_j,\label{casimir1}
\end{eqnarray}
with
\begin{eqnarray*}
\Omega_j&=&-\frac{\pi^2m_jg_{jj}\xi_j^2\widetilde{M}_j}{360\hbar^2\ell^3}.
\end{eqnarray*}
This quantity is called Casimir energy.

We now consider the Casimir force, which is defined as the first derivative of Casimir energy with respect to the distance
\begin{eqnarray}
F_C=-\frac{\partial\Omega}{\partial\ell}.\label{dinhnghia}
\end{eqnarray}
Combining (\ref{casimir1}) and (\ref{dinhnghia}) leads to
\begin{eqnarray}
F_C=\sum_{j=1,2}\left(-\frac{\pi^2m_jg_{jj}\xi_j^2\widetilde{M}_j}{120\hbar^2\ell^4}+\frac{\pi^2m_jg_{jj}\xi_j^2}{360\hbar^2\ell^3}\frac{\partial\widetilde{M}_j}{\partial\ell}\right).\label{force1}
\end{eqnarray}
\begin{figure}[htp]
  \includegraphics{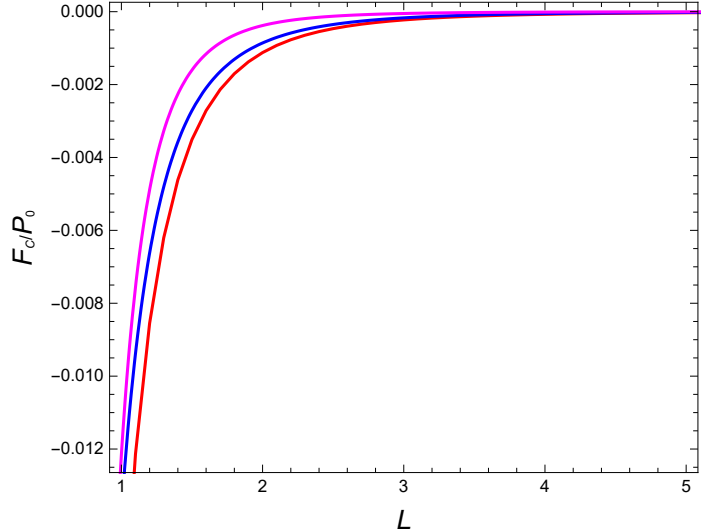}
  \caption{(Color online) The Casimir force versus $L$. The red and blue lines correspond to IHF and one-loop approximation at $K=3$, respectively. The magenta line associates strong segregated and IHF approximation.}\label{f3}
\end{figure}
The first thing we can say is that the same as in one-loop approximation \cite{Thu1}, the Casimir force is not simple superposition of the one of two single component BEC. In addition, It is very interesting to note that in IHF approximation the Casimir force differs from the one in one-loop approximation amount is last term in right hand side of Eq. (\ref{force1}). This gives several comments as follows:

- Firstly, the Casimir force is not proportional to $\ell^{-4}$ like the one in one-loop approximation.

- The second one, based on Fig.\ref{f1a} for $K=3$, we can conclude for general case that the first derivative of effective masses with respect to distance is negative, this fact leads to a result is that the strength of Casimir force in IHF approximation is larger than the one in one-loop approximation at the same value of other parameters.

- In addition, when $\ell$ is large enough the first derivative of effective masses with respect to distance is vanishing. In this region the same value for Casimir force is obtained in both IHF and one-loop approximation.

- The last but not least, the Casimir force is vanishing for ideal Bose gasses, i.e. $g_{jj}=0$. This result is the same as that in one-loop approximation.

\begin{figure}[htp]
  \includegraphics{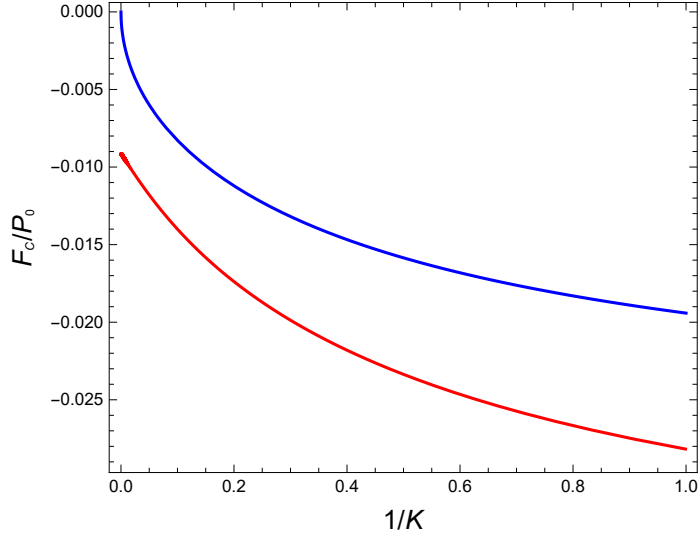}
  \caption{(Color online) The Casimir force versus $1/K$ at $L=1$. The red and blue lines correspond to IHF and one-loop approximation, respectively.}\label{f4}
\end{figure}

To illustrate for the above comments, the computation is made and shown in Fig. \ref{f3} with the same parameter in Figs. \ref{f1}. The red and blue lines correspond to the IHF and one-loop approximation. These lines confirm above comments. In case of strong segregated, from (\ref{strong1}) and (\ref{dinhnghia}) one has
\begin{eqnarray}
F_C=-\sum_{j=1,2}\frac{(m_jg_{jj}\pi^2)^{4/3}}{90.3^{2/3}.5^{1/3}\hbar^{8/3}\ell^5},\label{strong2}
\end{eqnarray}
and it is shown by magenta line in Fig. \ref{f3}.

The evolution of Casimir force versus $1/K$ is plotted in Fig. \ref{f4} at $L=1$ and other parameters are the same as in Figs. \ref{f2}. The red and blue lines correspond to IHF and one-loop approximation. This figure shows that the strength of Casimir force decreases as the interspecies interaction increases. This fact can be understandable if we note that the Casimir force is attractive whereas the interspecies interaction is impulsive. Eq. (\ref{strong2}) and Fig. \ref{f4} show that the Casimir force is non-zero in limit of strong segregated within IHF approximation, whereas it is vanishing in one-loop approximation.  This is an interesting result in comparing to the one in \cite{Thu1}. This result gives us the conclusion that the Casimir force is always on top of interspecies interaction and it is an improvement for our result in previous paper \cite{Thu1}.

\section{Conclusion and outlook\label{sec:4}}

In the foregoing sections, using quantum field theory in IHF approximation we consider the finite size effect in two component Bose-Einstein condensates. Many analytical calculations are worked out and numerical computations for the typical system of rubidium with two hyperfine states are also made. Our results show that the finite size effect produces significant changes on the static properties of BECs. Our main results are in order

- The effective masses and order parameters strongly depends on the distance $L$ between two slabs, this property can not be found if we consider within one-loop approximation. When $L$ is large enough these quantities approach to constants and coincide to those for infinite system.

- In IHF approximation we find the Casimir force is more accurate than that in on-loop approximation. In general, this force is not proportional to $\ell^{-4}$ as it is in one-loop approximation. However, when $\ell$ is large enough this rule is valid.

- We proved that the Casimir force is always on top of the interspecies interaction. This leads to nonzero value of Casimir force in strong segregated limit. This is our highlight result and improvement our result in previous work \cite{Thu1}.

It is very interesting if one can check these results by experiment.

\section*{Acknowledgements}

 This work is financial supported by the Vietnam National Foundation for Science and Technology Development (NAFOSTED) under Grant No.103.01-2018.02. The fruitful discussions with Shyamal Biswas are acknowledged with thanks.

\section*{References}

\bibliography{mybibfile}

\end{document}